\documentclass[%
 reprint,
 amsmath,amssymb,
 aps,
]{revtex4-2}

\usepackage{float}

\usepackage{xcolor}
\usepackage{amsmath}
\usepackage{amssymb}

\usepackage{graphicx}
\usepackage{dcolumn}
\usepackage{bm}
\usepackage{soul}

\begin{document}


\title{Spin-wave confinement in a hybrid superconductor-ferrimagnet nanostructure}

\author{Julia Kharlan$^{1,2}$, Krzysztof Sobucki$^{1}$, Krzysztof Szulc$^{1}$, Sara Memarzadeh$^{1}$, Jarosław W. Kłos$^{1,}$}
\email{klos@amu.edu.pl}
\affiliation{
$^{1}$ISQI, Faculty of Physics, Adam Mickiewicz University, Poznań, Poland\\
$^{2}$Institute of Magnetism NASU and MESU, Kyiv, Ukraine
}%

\date{\today}

\begin{abstract}{
Eddy currents in a superconductor shield the magnetic field in its interior and are responsible for the formation of a magnetic stray field outside of the superconducting structure. The stray field can be controlled by the external magnetic field and affect the magnetization dynamics in the magnetic system placed in its range. In the case of a hybrid system consisting of a superconducting strip placed over a magnetic layer, we predict theoretically the confinement of spin waves in the well of the static stray field. The number of bound states and their frequencies can be controlled by an external magnetic field. We have presented the results of semi-analytical calculations complemented by numerical modeling.}

\end{abstract}

\maketitle


\section{Introduction \label{sec:Intro}}
The states of superconductivity and ferromagnetism are very rarely observed in a single material. Their intrinsic coexistence (i.e., in one uniform phase, for the same electrons) was found for triplet pairing in a proximity to a magnetic quantum critical point (e.g. for ${\rm UGe_2}$) \cite{ran_nearly_2019,saxena_superconductivity_2000}. The other possibility, known for a long time and more conventional \cite{Sinha1982,Bernhard1999}, is the coexistence of two phases where large and localized moments of 4f electrons (Er, Gd) provide long-range strong ferromagnetism whereas 3d conduction electrons are responsible for superconductivity.

The hybrid systems \cite{Lyuksyutov2005,linder2015,Bespalov2022,Cai2023}, where the superconductor and ferromagnet are part of the same structure and interact with each other, usually offer much more flexibility both in the design and implementation of new features. The superconductor/ferromagnet hybrids can be divided into two categories -- where both subsystems are in direct contact \cite{Bergeret_2001} or separated by a nonmagnetic, nonconducting material. In the latter case, the coupling at a distance results from the fact that both the eddy currents in the superconductor and the magnetic moments in the ferromagnet generate a magnetic field. The coupling provided by the magnetic field can be controlled by the external magnetic field and can be tailored by the geometry, since both the distribution of the eddy currents and the magnetization configuration depend on these factors.

In electromagnetically-coupled hybrids, the ferromagnet can modify the properties of the superconductor, e.g., the magnetic screening can increase the value of critical current density in the superconductor \cite{Genenko1999}, or the presence of the stray field produced by ferromagnetic nanoelements can affect the nucleation of vortices \cite{Milosevic2005} or pin and guide the vortices in the superconductor \cite{Vlasko2017}. Similarly, the presence of the superconductor can have influence on the ferromagnet, e.g. by controlling the magnetization dynamics {\cite{Golovchanskiy2018, Borst_2032}. In this research field, we can find the reports about induction of magnonic crystals and nonreciprocal spin-wave (SW) transmission in uniform magnetic layer due to the screening of the dynamic demagnetizing field by a superconductor {\cite{Golovchanskiy2019,Kuznetsov2022}, magnon--phonon interaction \cite{Ghirri_2023, ghirri2023interplay}, the gating of magnons induced by superconducting current \cite{Yu_2022, Zhou_2023}, enhancement of nonlinear SW dynamics \cite{Golovchanskiy2020}, Bragg scattering of SWs on the field produced by the Abrikosov vortex lattice \cite{Dobrovolskiy2019,Niedzielski2023}, or SW generation by moving vortices \cite{Shekhter2011, Dobrovolskiy_2022, dobrovolskiy2023}. Undoubtedly, superconductor/ferromagnet hybrids offer many possibilities for controlling the dynamics of SWs. One of the topics not fully explored is the problem of localization of SWs in these systems.

In this study, we conduct a theoretical and numerical investigation of the SW confinement induced in a uniform ferrimagnetic (FM) layer by the stray field of a superconducting (SC) strip. We demonstrate that by adjusting the applied field, we can impact on the depth of the stray field well and thereby control the number and frequencies of the confined SW modes. 

\section{Model \label{sec:Model}}

The considered hybrid system consists of a FM gallium-doped yttrium iron garnet (Ga:YIG) thin film and a SC Nb strip in the Meissner state, electrically isolated from each other by thin nonmagnetic spacer (see Fig.~\ref{fig:structure}). According to the Meissner effect, a SC strip expels a magnetic field from its volume by means of eddy currents. These currents create a nonuniform distribution of the magnetic field throughout the entire space, including the FM film. Such a geometry enables the investigation of the coupling between FM and SC subsystems as purely classical, where the stray magnetic field from the eddy currents in the SC strip impacts the magnetization in the FM layer.

\begin{figure}
\includegraphics[width=0.95\columnwidth]{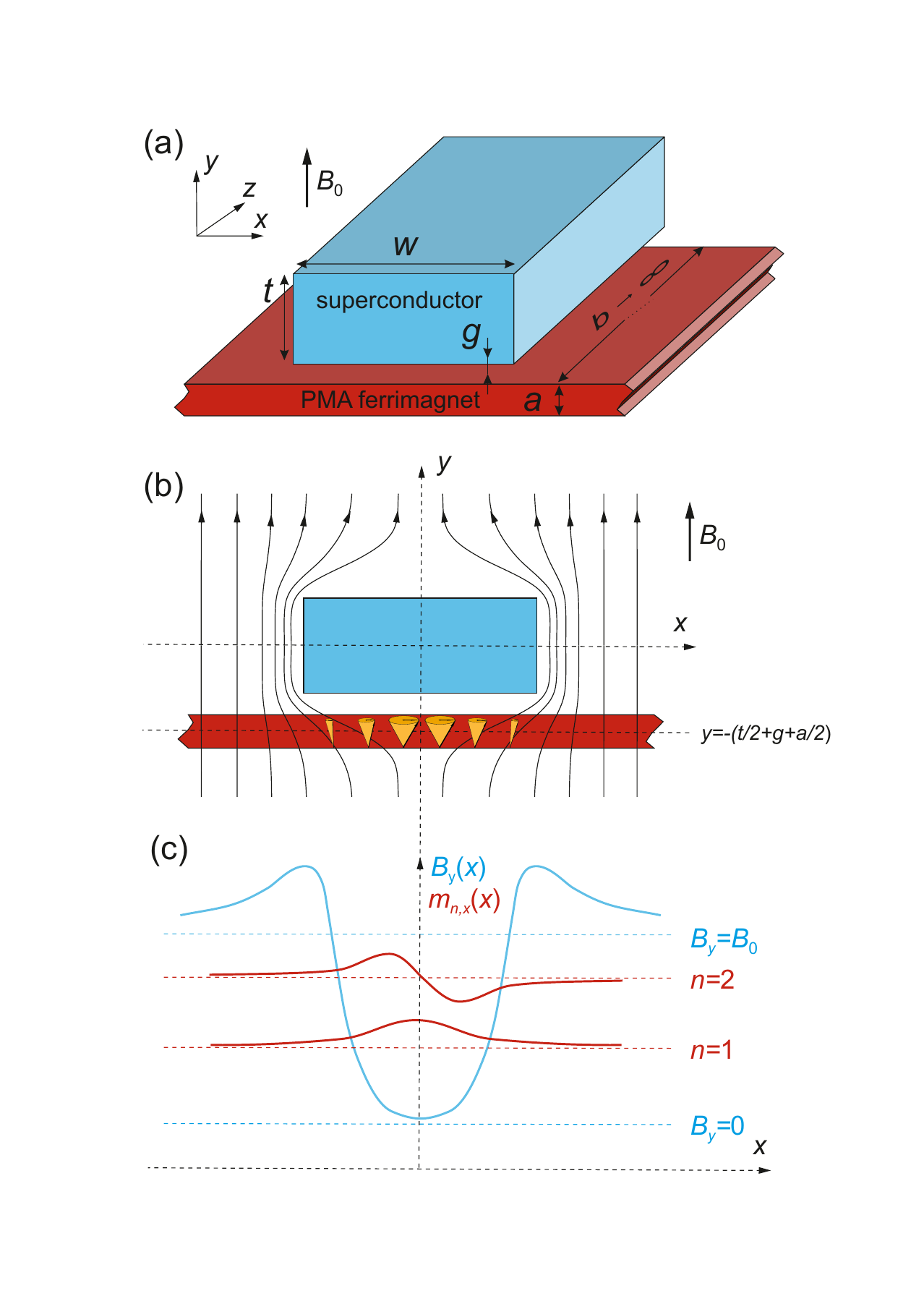}
\caption{(a) A thin FM film ($a=20$ nm) is exposed to the stray field of a rectangular SC strip ($w=400$ nm, $t=100$ nm). The ferrimagnet (Ga:YIG) and the superconductor (Nb) are separated by a small gap ($g=10$ nm). The magnetic film has low $M_{\rm s}$ and exhibits PMA, so it is magnetized out of plane in the absence of the external magnetic field. Sketches (b) and (c) illustrate the mechanism of SW localization. (b) The internal static magnetic field is lowered in the region of the ferrimagnet underneath the SC strip. This leads to the confinement of SW modes (c), which are quantized in the quasiparabolic well of the internal field.}
\label{fig:structure}
\end{figure}

The system is placed in an external magnetic field perpendicular to the FM layer. In Ga:YIG, the shape anisotropy is overcome by the perpendicular magnetic anisotropy (PMA), leading to the magnetization being directed out of plane even in the absence of an external magnetic field. Operating with the relatively small external field, we can sustain the Meissner state in a confined geometry of the strip, and observe the impact of the stray field of the SC strip on the magnetization dynamics in the FM layer. Then, the stray field of SC strip is tuned by the external field and induces the well of the static effective field of controllable depth in the FM layer. The well can confine the SWs of  frequencies lower than the ferromagnetic resonance (FMR) frequency of the FM layer in the absence of a SC strip. 
 
Any static stray field is not produced outside a uniformly magnetized infinite FM layer -- only the internal demagnetizing field ($-M_{\rm s}\hat{\mathbf{y}}$) is present. However, the stray field of the SC strip induces a weak magnetization texture in the FM layer, close to the SC-strip edges. It can produce a static stray field, although it is small and can be neglected -- see Appendix~\ref{app:micromag}.
The dynamic stray field coming from the SW modes in the FM layer can also be  considered  negligible. For the FM layer magnetized in the out-of-plane direction, the dynamic components of magnetization are tangential to the surfaces and do not produce any surface magnetic charges. On the other hand, the dynamic stray field can be generated by the volume magnetic charges, but it requires a strong nonuniformity of the SW profile, which is observed for the higher SW modes (not investigated here). Therefore, we assume that eddy currents are unaffected by SWs and remain constant. To sum up, both static and dynamic stray fields induced in the FM layer are negligible for the SC strip. Therefore, there is no need to solve the self-consistent problem for our system and take into account the mutual interaction between the FM layer and the SC strip. For the considered magnetization configuration in the FM layer (i.e. out-of-plane saturation), the system is in forward volume geometry so any change in the dynamic stray field resulting from the presence of the SC strip will not induce the nonreciprocity effect observed in superconductor/ferromagnet hybrids \cite{Kuznetsov2022}. Therefore, the time-reversal symmetry remains unbroken and the confined SWs have a form of standing modes.

Taking these assumptions into account, our studies were carried out in two stages. We first calculated the static stray field generated by the SC strip. It was determined from the distribution of SC currents, which was found by a semi-analytical solution of the London equations. The static field generated by the SC strip was then included as a correction to the effective field of the Landau--Lifshitz (LL) equation. The LL equation was solved both semi-analytically and numerically and was used to find the confined SW modes.

\subsection{Static magnetic field produced by a superconducting strip}

The London equation has been solved analytically for a number of geometries, such as a film \cite{Schmidt1997}, a cylindrical wire \cite{Fiolhais2014}, and an infinitely-thin cylindrical dot \cite{Badia1998}. However, the analytical solution of the London equation presents difficulties, even for such a simple structure as a strip, due to the impossibility of variable separation. We have therefore used the semi-analytical method developed by Brandt \cite{Brandt1994a,Brandt1994b}. For the SC strip in an external field applied along the $y$-direction $\mathbf{B}_0=B_0\hat{\mathbf{y}}$, the Meissner effect induces a current of density $\mathbf{J}=J(x,y)\hat{\mathbf{z}}$ flowing along the $z$-axis and generating a stray field $\mathbf{B}_{\rm sc}(x,y)$, which lacks $z$-component -- see Fig.~\ref{fig:structure}(b). According to  Amp\`ere's law, we can relate the current density to the total field $\mathbf{B}$ and vector potential $\mathbf{A}$ as
\begin{equation}
    \mu_0\mathbf{J} = \nabla\times\mathbf{B}=\nabla\times\nabla\times\mathbf{A}=-\Delta\mathbf{A},\label{eq:amper}
\end{equation}
where we have used the Coulomb gauge ($\nabla\cdot \mathbf{A}=0$), and the symbol $\mu_0$ denotes the permeability of vacuum. Eq.~(\ref{eq:amper}) can be interpreted as a two-dimensional Poisson equation for vector potential $\mathbf{A}=A(x,y)\hat{\mathbf{z}}$ with a current $\mathbf{J}=J(x,y)\hat{\mathbf{z}}$ being a source:
\begin{equation}
    \Delta A(x,y)=-\mu_0 J(x,y),\label{eq:poisson}
\end{equation}
where $\Delta$ is a two-dimensional Laplacian. Let us decompose the  vector potential $\mathbf{A}=(x B_0 +A_{\rm sc}(x,y))\hat{\mathbf{z}}$ into the contribution related to the uniform external field: $\mu_0 x H_0$ and the field produced by SC, $A_{\rm sc}(x,y)$, and then relate the latter inside the SC to a current $J(x,y)$ using the London equation with Coulomb gauge:
\begin{equation}
    A_{\rm sc}(x,y)=-\mu_0\lambda^2 J(x,y),\label{eq:london}
\end{equation}
with $\lambda$ the London penetration depth. Then, using Eq.~(\ref{eq:london}), we can write Eq.~(\ref{eq:poisson}) in the following integral form:
\begin{equation}
    \lambda^2 J(x,y)\!=\!\!\!\iint_{S} dx'dy' Q(x,y,x',y')J(x',y')+\!x H_0,\label{eq:curr_int}
\end{equation}
where $S$ is the cross-section of the SC strip in $xy$-plane and the function $Q=\tfrac{1}{2\pi}\ln(\sqrt{(x-x')^2+(y-y')^2})$ is the integral kernel for two-dimensional Laplace operator.

Equation~(\ref{eq:curr_int}) can be integrated numerically by sampling the function $J(x,y)$ on a square grid of equidistant points $(x_i,y_j)$: $x_i=((i-\tfrac{1}{2})/N_x-\tfrac{1}{2})w$, $y_j=((j-\tfrac{1}{2})/N_y-\tfrac{1}{2})t$, where $w$ and $t$ are the strip width and thickness respectively (see Fig.~\ref{fig:structure}) and $N_x$ and $N_y$ are number of points in the $x$- and $y$-directions, where $N_x/w\approx N_y/t$ and $i=1,\ldots,N_x$, $j=1,\ldots,N_y$. By marking the points $\mathbf{r}_l=(x_i,y_j)$ with an index $l=1,\ldots,N\!=\!N_xN_y$, the function $J(x,y)$ becomes a vector $J_l$ with $N$ components, and the integral kernel is transformed into the $N\times N$ matrix $Q_{l,l'}$. Note that $Q_{l,l'}$ diverges for $l=l'$, which should be avoided. Equation~(\ref{eq:curr_int}) for the distribution of the current density $J(x,y)$ can then be rewritten in  matrix form:
\begin{equation}
    \lambda^2 J_l=\sum_{l'}\tilde{Q}_{l,l'}J_{l'}+\frac{1}{\mu_0}x_{l'}B_0,\label{eq:curr_mat}
\end{equation}
where $\tilde{Q}_{l,l'}=Q_{l,l'}\frac{w}{N_x}\frac{t}{N_y}$. The solution of Eq.~(\ref{eq:curr_mat}) for $J_l$ can be found when we invert the matrix $P_{l,l'}=\lambda^2 \delta_{l,l'} -\tilde{Q}_{l,l'}$:
\begin{equation}
    J_l=\frac{1}{\mu_0}B_0\sum_{l'}(P^{-1})_{l,l'}x_{l'}.\label{eq:curr_elem}
\end{equation}

Finally, the distribution of the field $B_{\rm sc}$ produced by eddy currents can be determined using the Biot--Savart law:
\begin{equation}
    \begin{split}
    B_{{\rm sc},x}(x,y)=\frac{\mu_0}{2\pi}\iint_{S} dx'dy'\frac{J(x',y')(y'-y)}{(x'-x)^2+(y'-y)^2},\\
    B_{{\rm sc},y}(x,y)=-\frac{\mu_0}{2\pi}\iint_{S} dx'dy'\frac{J(x',y')(x'-x)}{(x'-x)^2+(y'-y)^2}.\label{eq:b_field}
    \end{split}  
\end{equation}
The integrals in Eq.~(\ref{eq:b_field}) can be calculated numerically by sampling the current density $J$ on the cross-section $S$ of the SC strip.

\subsection{Spin-wave modes confined in the well of the magnetic field induced by a superconducting strip}

After calculating the stray field generated by the SC strip, the next step is to investigate the SW localization in the FM layer placed under the strip. Two assumptions were made in our theoretical model: (i) our calculations have shown that the magnetic field generated by the SC strip does not vary significantly across the thin FM layer, so we assume that the SC field does not depend on the $y$-coordinate and is equal to the value at the film center; (ii) we have neglected the tangential component of the SC field and, consequently, the deviation of the magnetization from the film normal, so we have included only the normal component of the SC field, which is responsible for the formation of the well for localized SW modes, and thus considered the uniform magnetization directed perpendicular to the film plane.  

The magnetization dynamics in the continuous medium is described semi-classically by the LL equation:  
\begin{equation}
    \frac{\partial \mathbf{M}(\mathbf{r},t)}{\partial t}=-|\gamma|\mu_0 \mathbf{M}(\mathbf{r},t)\times\mathbf{H}_{\rm eff}(\mathbf{r},t). \label{eq:LL}
\end{equation}
Here  $\gamma$ is the gyromagnetic ratio and $\mathbf{H}_{\rm eff}$ is the effective magnetic field, calculated from the free energy density \cite{Gurevich1996} 
\begin{equation}
  \begin{split}
    F = -(\mathbf{B}_0 + \mathbf{B}_{\rm sc}) \cdot \mathbf{M} + \frac{A_{\rm ex}}{M_{\rm s}^2} (\nabla\mathbf{M})^2 \\ - \frac{1}{2}\mu_0\mathbf{H}_{\rm d} \cdot \mathbf{M} - \frac{K_{\rm u}}{M_{\rm s}^2}(\mathbf{M} \cdot \hat{\mathbf{n}})^2\label{eq:F}
  \end{split}
\end{equation}
with respect to magnetization: $\mathbf{H}_{\rm eff}=-\frac{1}{\mu_0}\frac{\delta F}{\delta \mathbf{M}}$. Also, in Eq.~(\ref{eq:LL}), we neglected the damping. The material parameters in Eq.~(\ref{eq:F}): $M_{\rm s}$, $A_{\rm ex}$ and $K_{\rm u}$ denote the saturation magnetization, exchange-stiffness constant and uniaxial anisotropy, respectively. In our case, the magnetocrystalline anisotropy is easy-axis anisotropy ($K_{\rm u}>0$), oriented in the out-of-plane direction ($\hat{\mathbf{n}}=\hat{\mathbf{y}}$) that overcome the shape anisotropy ($K_{\rm u}>\mu_0M^{2}_{\rm s}/2$). The demagnetizing field is nonlocal which means that $\mathbf{H}_{\rm d}(\mathbf{r},t)$ at specific point $\mathbf{r}$ depends on the magnetization distribution in the whole magnetic body. We use the linear approximation, where the magnetization precesses harmonically around the equilibrium position $\mathbf{M}_0(\mathbf{r})=M_{\rm s}\hat{\mathbf{y}}$ with  amplitude $\mathbf{m}(\mathbf{r})=m_x\hat{\mathbf{x}}+m_z\hat{\mathbf{z}}$ much smaller than the static magnetization component $|\mathbf{m}(\mathbf{r})|\ll M_{\rm s}$, giving the total magnetization vector $\mathbf{M}(\mathbf{r},t)\approx\mathbf{M}_0(\mathbf{r})+\mathbf{m}(\mathbf{r})e^{i\omega t}$. Then, we can decompose the demagnetizing field into a static and dynamic component $\mathbf{H}_{\rm d}(\mathbf{r},t)\approx -M_{\rm s}\hat{\mathbf{y}}+\mathbf{h}_{\rm d}(\mathbf{r})e^{i\omega t}$, where the dynamic component $\mathbf{h}_{\rm d}=h_{\mathrm{d},x}\hat{\mathbf{x}}+h_{\mathrm{d},z}\hat{\mathbf{z}}$ can be written in  general form: 
\begin{equation}
    \mathbf{h}_{\rm d}(\mathbf{r})=-\nabla\int_{V}dv'\mathbf{m}(\mathbf{r}')\cdot\nabla'\frac{1}{4\pi|\mathbf{r}-\mathbf{r}'|}.\label{eq:dyn_demag}
\end{equation}
The LL equation (\ref{eq:LL}) can then be  written in the linearized form:
\begin{equation}
   \begin{split}
   \;\;i\omega\; m_x(x)=&|\gamma|\left[\left(B_0-\mu_0M_{\rm s}+\frac{2 K_u}{M_{\rm s}}+B_{{\rm sc},y}(x)\right)m_z(x)\right.\\&-\left.\frac{2 A_{\rm ex}}{ M_{\rm s}}\frac{d^2}{dx^2} m_z(x)-\mu_0 M_{\rm s}h_{{\rm d},z}(x)\right],\\
   -i\omega\; m_z(x)=&|\gamma|\left[\left(B_0-\mu_0M_{\rm s}+\frac{2 K_u}{M_{\rm s}}+B_{{\rm sc},y}(x)\right)m_x(x)\right.\\&-\left.\frac{2 A_{\rm ex}}{M_{\rm s}}\frac{d^2}{dx^2} m_x(x)-\mu_0 M_{\rm s} h_{{\rm d},x}(x)\right].
   \end{split}\label{eq:lin_LL}
\end{equation}
Here we have used the assumption that inside the thin FM layer both the magnetization and field are constant in an out-of-plane direction, i.e., they do not depend on the $y$-coordinate, and the tangential component of the stray field of the SC can be neglected: $B_{{\rm sc},x}\approx~0$.

In most cases, it is impossible to obtain a rigorous analytical solution of the LL equation for the SWs in nanostructures. Usually, the eigenfunctions of the exchange-field operator
, which, however, are not eigenfunctions of the operator for the demagnetizing field, are used as trial functions in the Ritz method. Such an approach, when applied to planar nanoelements with relatively uniform magnetization distribution, allows one to obtain the solutions that are close to the exact ones  \cite{Kakazei2004,Kakazei2012,Bunyaev2015,Klein2008}. In these systems, the demagnetizing field is almost constant in the whole volume (due to the small magnetic volume charges), except in the vicinity of the edges (where surface magnetic charges play role). However, such an approach does not work in the case of nanoelements with strongly inhomogeneous demagnetizing or external fields. In such cases, it is necessary to consider trial solutions being the eigenfunctions of the operator, which is a sum of the exchange operator and the operators that express the impact of demagnetizing effects or external fields. For instance, the SW localization in a tangentially magnetized thin layer induced by the dipolar stray field of the magnetic sphere was considered in \cite{Tartakovskaya2016}. In this system, the static stray field of the sphere, perceptible by the SW in the FM layer, was approximated by parabolic functions. In the framework of this approximation, the linearized LL equation was reduced to a system of coupled Schr\"odinger-like equations for a quantum harmonic oscillator with the dynamic demagnetizing field of the FM  layer considered as a perturbation \cite{Tartakovskaya2016,Kharlan2019,Kalinikos1986}. Therefore, the eigenfunctions in the form of Hermite polynomials could be chosen as trial solutions for SW modes. 

Considering the assumptions made at the beginning of Sec.~\ref{sec:Model}, we can use the approach proposed in Ref.~\cite{Tartakovskaya2016} for investigating superconductor--ferrimagnet hybrid system. The linearized LL equation can then be  written in a compact form:
\begin{equation}
     \begin{split}
   \;\;i\omega\; m_x(x)&=|\gamma|\big( \Xi(x) m_z(x)-\mu_0 M_{\rm s}h_{{\rm d},z}(x)\big),\\
   -i\omega\; m_z(x)&=|\gamma|\big( \Xi(x) m_x(x)-\mu_0 M_{\rm s}h_{{\rm d},x}(x)\big).
   \end{split}\label{eq:lin_LL_op}  
\end{equation}
The operator $\Xi(x) = -\tfrac{2A_{\rm ex}}{M_{\rm s}} \tfrac{d^2}{dx^2} + B_{y}(x)$ describes the impact of dynamic exchange field and total static field:
\begin{equation}
    B_{y}(x)=\! B_0 -\mu_0M_{\rm s}+\frac{2K_{\rm u}}{M_{\rm s}} + \underbrace{B_0(\beta_0+\beta_2 x^2 + O(x^2))}_{B_{{\rm sc},y}(x)},\label{eq:B_expansion}
\end{equation}
on magnetization dynamics. It is worth noting that, according to Eqs.~(\ref{eq:curr_elem}) and (\ref{eq:b_field}), the stray field $B_{{\rm sc},y}(x)$ is linearly scaled by external field $B_0$.

By introducing the circular polarization for the dynamic component of  magnetization, $m=m_z + i m_x$, and neglecting the dynamic demagnetizing field $\mathbf{h}_{\rm d}$, we can transform the set of equations (\ref{eq:lin_LL_op}) into a form equivalent to the stationary Schr\"{o}dinger equation:
\begin{equation}
    \Xi(x)\,m_{n}(x)=\frac{1}{|\gamma|}\omega_{0,n} m_{n}(x).\label{eq:Schrodinger}
\end{equation}
Equation (\ref{eq:Schrodinger}) can be solved analytically after neglecting higher-order terms in the expansion in Eq.~(\ref{eq:B_expansion}). The problem is then reduced to the magnonic counterpart of the quantum harmonic oscillator with equidistant  eigenfrequencies $\omega_{0,n}$ and corresponding eigenfunctions $m_{n}(x)$ expressed by Hermite polynomials $H_n(\xi)~=~(-1)^{n}\exp(\xi^2)\tfrac{d^n}{d\xi^n}\exp(-\xi^2)$:
\begin{equation}
  m_n(\xi=Kx)=C M_{\rm s}\underbrace{\frac{1}{\sqrt{2^n n!\sqrt{\pi}}}\exp(-\xi^2)H_n(\xi)}_{\psi_n(\xi)}\label{eq:psi}
\end{equation}
for every $n^{\rm th}$ SW mode. The scaling factor $K=\sqrt[4]{B_0 \, \beta_2 \, \tfrac{ M_{\rm s}}{2A_{\rm ex}}}$ expresses the interplay between the strength of the exchange interaction and the curvature (strength) of the well of the SC stray field, in which the SW modes are confined. The real function $\psi_n(\xi)$ is normalized, $\int_{-\infty}^{\infty}\psi_n(\xi)=1$, and the dimensionless constant $C$ expresses the complex SW amplitude. 
 
The approximation $B_{{\rm sc},y}(x)\approx B_0(\beta_0+\beta_2 x^2)$ is not valid for wide SC strips, $w\gg\lambda$, (where the well $B_{{\rm sc},y}(x)$ becomes relatively flat at the bottom) and narrow strips, $w<\lambda$, or small $B_0$ (where the well is shallow and distortion of its parabolic shape, related to the presence of the barriers on its sides, affects the mode of the lowest frequency). 
Therefore, we solve Eq.~(\ref{eq:Schrodinger}) numerically for the stray field $B_{{\rm sc},y}$ in its full form given by Eq.~(\ref{eq:b_field}) -- this task is relatively simple and does not demand extensive computations. The obtained frequencies $\tilde{\omega}_{0,n}\ne{\omega}_{0,n}$ of the bound SW modes were calculated for a realistic (i.e., nonparabolic) stray field of the SC strip but with dynamic dipolar interaction neglected. However, we used the approximated (by Hermite polynomials) profiles of the eigenmodes from Eq.~(\ref{eq:psi}) to include the role of dynamic dipolar interactions.

The next step in our computational procedure is to include the impact of the dynamic demagnetizing field $\mathbf{h}_{\rm d}$ and correct the frequencies of individual modes $\tilde{\omega}_{0,n}$ calculated in the absence of $\mathbf{h}_{\rm d}$. We are going to limit our calculations to the so-called diagonal approximation \cite{Kalinikos1986}, which was satisfactory for many similar cases \cite{Kharlan2019,Tartakovskaya2016}, where the difference between successive frequencies $\tilde{\omega}_{0,n}$ is large with respect to the frequency shift due to the dynamic dipolar interaction. This approximation also neglects  the intermode dipolar coupling. According to Eq.~(\ref{eq:dyn_demag}), the dynamic demagnetizing field produced by a single mode can be written in the form
\begin{equation}
    h_{{\rm d},p,n}(x)=-\int_V dv' \sum_{q} m_{q,n}(x') \frac{d^2}{dp dq'}\frac{1}{4\pi |\mathbf{r}-\mathbf{r}'|},\label{eq:hd_gen}
\end{equation}
where the indices $p$ and $q$ run over the coordinates $x$ and $z$, and the derivative $\frac{d}{dq'}$ means $\frac{d}{dx'}$ (or $\frac{d}{dz'}$) for $q=x$ (or $q=z$). We denote by $m_{q,n}(x)$  the components of dynamic magnetization for the $n^{\rm th}$ mode. 

The spectrum of eigenmodes, which includes the correction resulting from the presence of dynamic dipolar interaction, has the  form (see Appendix~\ref{app:disp} for a derivation):
\begin{equation}
    \omega_n\!=\!\sqrt{\big(\tilde{\omega}_{0,n}\!-\!|\gamma|\mu_0 M_{\rm s}	\langle \tilde{h}_{{\rm d},x}	\rangle_n\big)\big(\tilde{\omega}_{0,n}\!-\!|\gamma|\mu_0 M_{\rm s}	\langle \tilde{h}_{{\rm d},z}\rangle_n\big)},\label{eq:omega_diag}
\end{equation}
where the $\langle \tilde{h}_{{\rm d},p}\rangle_n$ are averaged and normalized (dimensionless) components of $\mathbf{h}_{\rm d}$. The averaged field $\langle \tilde{h}_{{\rm d},z}\rangle_n$ is equal to zero in the considered system, whereas $\langle \tilde{h}_{{\rm d},x}\rangle_n$ can be expressed by the compact formula (see Ref.~\cite{Tartakovskaya2016} and Appendix~\ref{app:avg_hd}):
\begin{equation}
  \langle\tilde{h}_{{\rm d},x}\rangle_n\!\approx-\frac{1}{K}\int_{-\infty}^{\infty}\!dk_x\left(1\!-\!\frac{1-e^{-k_x a}}{k_x a}\right)\tilde{\psi}_{n}\left(\frac{k_x}{K}\right),\label{eq:h_dyn_aver}
\end{equation}
where $\tilde{\psi}_{n}(\mathcal{K})=\tfrac{1}{\sqrt{2\pi}}\int_{-\infty}^{\infty}e^{-i\mathcal{K}\xi}\psi_{n}(\xi)d\xi$ is the Fourier transform of $\psi(\xi)$. The one-dimensional integral in Eq.~(\ref{eq:h_dyn_aver}) is relatively simple to compute  compared to the calculations of $\langle \tilde{h}_{{\rm d},u}\rangle_n$ directly from Eq.~(\ref{eq:hd_gen}). It is worth noting that we approximate $m_{p,n}(x)$ by taking $m_{p,n}(x)\approx C_p M_{\rm s}\psi_n(Kx)$, where the $\psi_n(Kx)$ are eigenfunctions of $\Xi(x)$. The amplitudes $C_p$ are canceled during  normalization -- see Appendix~\ref{app:avg_hd}.

\subsubsection*{Micromagnetic simulations}

To perform micromagnetic simulations, we used the open-source environment Mumax3 \cite{vansteenkiste2014design} that solves the LL equation using the finite-difference method in the time domain. We discretized the system with a regular mesh with unit cell $2\times 20 \times 20$~nm$^3$ (along the $x$-, $y$-, and $z$-axes respectively). The system was placed in an inhomogeneous external magnetic field $\mathbf{B}_0+\mathbf{B}_{\rm sc}(\mathbf{r})$ that includes the full contribution of the stray field of the SC strip. The field profile was imported from an external file containing the results of semi-analytic calculations of $\mathbf{B}_{\rm sc}(\mathbf{r})$. We applied the periodic boundary conditions along the $z$-axis (1024 repetitions) to simulate an infinitely-long strip. We increased the damping constant close to the edges of the domain perpendicular to $x$-axis from $\alpha_\mathrm{edge}=0$ to $0.5$ in order to eliminate the reflections along $x$-axis edges. The SWs in the system were numerically excited by a rectangular model-antenna of width $12$~nm, placed in the center of the film. The antenna generated a local dynamic magnetic field described by the function $H_\mathrm{ant}(t)=H_{0}\,\mathrm{sinc}(2\mathrm{\pi} f_\mathrm{cut}(t-t_{0}))\hat{x}$, where $\mu_{0}H_{0}=0.024B_{0}$ and $f_\mathrm{cut}$ is a cutoff frequency that varies in the simulations depending on the depth of the well of the stray field  $\Delta B_{{\rm sc},y}$ -- see Fig.~\ref{fig:sc_stray_field}(a). For each $\Delta B_{{\rm sc},y}$, the simulation was run for a time span of $1000/f_\mathrm{cut}$.
The results of the simulations were saved continuously as  snapshots of the magnetic configuration with time sampling $t_\mathrm{sample}=0.5/(1.5f_\mathrm{cut})$. In the postprocessing, all saved magnetic configurations were used to calculate the spectra by employing the Fourier transform. The visualizations of particular modes in the system were done by calculating the inverse Fourier transform of frequencies corresponding to the modes.  

Micromagnetic simulations were also used to calculate the stray field produced by magnetization texture in the FM layer resulting from the action of a stray field from the SC strip -- see Appendix~\ref{app:micromag}.

\subsubsection*{Finite-element method}
Numerical simulations were performed in COMSOL Multiphysics using the finite-element method in order to calculate the stray field coming from the SC strip subjected to the external magnetic field. To be consistent with the semi-analytical calculations, we used the London equation\begin{equation}
    \Delta A = \frac{1}{\lambda^2} A
\end{equation}
inside the SC material and Ampere's law
\begin{equation}
    \Delta A = 0
\end{equation}
outside of the SC material, both modified to the form for magnetic vector potential $A$. We investigated the system in the two-dimensional model in the $x-y$ plane. In this case, the Coulomb gauge $\nabla \cdot \mathbf{A} = 0$ and the condition on the SC boundary $\mathbf{A} \cdot \hat{\mathbf{n}} = 0$ are automatically fulfilled. The external magnetic field $B_0$ was applied as a boundary condition $-\partial_x A=B_0$ far away from the SC material. Here, the SC strip was placed in the center of a 100 $\mu$m square-shaped cell, on which boundary this condition is applied. 

We  checked the validity of the London model for the considered SC structure by the finite-element method calculation of the stray field based on the Ginzburg--Landau model -- see Appendix~\ref{app:stray_field_GL}.

\section{\label{sec:Res}Results}
We considered an SC strip of  thickness $t = 100$~nm and width $w = 400$ (or $800$~nm, for comparison). The strip was made of Nb, for which we took the London penetration depth $\lambda = 50$~nm. The assumed value is close to the experimental one determined for bulk Nb (47 nm) \cite{Maxfield1965}, but in the layer $\lambda$ is increased \cite{Gubin2005}. However, these changes are not significant for layers thicker than 100 nm. The SC strip was placed over the FM layer of thickness $a = 20$~nm and separated from its top surface by the nonmagnetic, nonconducting spacer of thickness $g=10$~nm -- see Fig.~\ref{fig:structure}. The FM layer, made of Ga:YIG, was characterized by the following values of material parameters: saturation magnetization $M_{\rm s}=16$~kA/m, exchange stiffness $A_\mathrm{ex}=1.37$~pJ/m and gyromagnetic ratio $\gamma=179$~GHz/T. The Ga:YIG layer can have a strong PMA \cite{Bottcher2022}. We assumed the realistic value of the uniaxial anisotropy $K_\mathrm{u}=756$~J/$\mathrm{m}^3$, which ensures the out-of-plane orientation of magnetization even in the absence of an external magnetic field. For micromagnetic simulations, we included the magnetization damping constant for Ga:YIG $\alpha=0.001$. For semi-analytical calculations the damping was neglected.

\subsection{Stray field of the superconducting strip}

\begin{figure}[b]
\centering
\includegraphics[width=0.9\columnwidth]{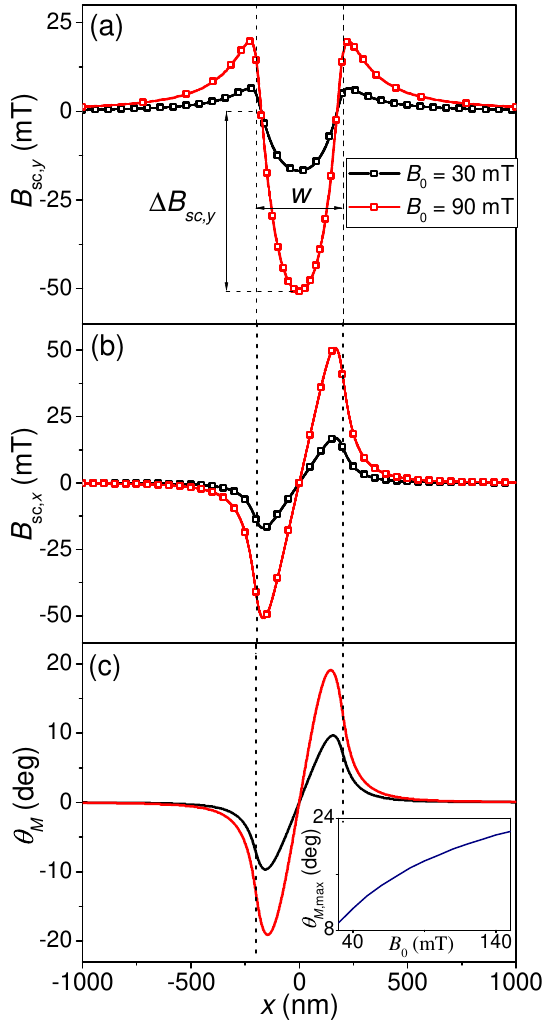}
\caption{The profiles of the magnetic field components (a) $B_{{\rm sc},y}$ and (b) $B_{{\rm sc},x}$ produced by SC strip, calculated using Eq.~(\ref{eq:b_field}) (solid lines) and using numerical simulations (open squares). (c) The deflection $\theta_M$ of the magnetization from the direction of the external field $\mathbf{B}_0=B_0\hat{\mathbf{y}}$. The inset plot shows the evolution of the maximum angle deviation with the external magnetic field value. Profiles (a)-(c) are plotted along the line $y=-70$~nm, passing through the center of the film. Dashed vertical lines mark the edges of the SC strip. Black and red lines correspond to the external field $B_0$ equal to 30 mT and 90 mT, respectively. Note that to calculate the total magnetic field, we have to sum $\mathbf{B}_0$ and $\mathbf{B}_{{\rm sc}}$.
}\label{fig:sc_stray_field}
\end{figure} 

Figure~\ref{fig:sc_stray_field}(a,b) present the profiles of the $y$-component (Fig.~\ref{fig:sc_stray_field}(a)) and $x$-component (Fig.~\ref{fig:sc_stray_field}(b)) of the stray field $\mathbf{B}_{\rm sc}$ produced by SC strip of  width $w=400$~nm for two selected values of out-of-plane applied field: $B_0=30$~mT (black lines) and $90$~mT (red lines). The profiles are the $x$-coordinate dependencies calculated for fixed $y=y_0=-(t/2+g+a/2)=-70$ nm, i.e. in the middle of the FM film. The dependencies $B_{{\rm sc},y}(x)$, $B_{{\rm sc},x}(x)$ are consistent with the expected onionlike shape of the field around the SC strip (see Fig.~\ref{fig:structure}(b)). The well of $B_{{\rm sc},y}(x)$ is a signature of the screening of the external field $\mathbf{B}_{0}=B_0\hat{\mathbf{y}}$ by the cost of the increase of the field close to the edges of the strip (vertical dashed lines in Fig.~\ref{fig:sc_stray_field}). On the other hand, the in-plane component of the stray field $B_{{\rm sc},x}$ reflects the deflection of the magnetic field lines by-passing the SC strip. This effect is the strongest in the proximity of the strip edges. The stray field was calculated semi-analytically using Eq.~(\ref{eq:b_field}) -- (red and black lines in Fig.~\ref{fig:sc_stray_field}(a,b)), and the obtained results were successfully cross-checked using finite-element method calculations -- (red and black open squares in Fig.~\ref{fig:sc_stray_field}(a,b)). Equations~(\ref{eq:b_field}) and the plots in Figs.~\ref{fig:sc_stray_field}(a,b) show that the stray field is linearly scalable with the external field $B_0$. While we increase $B_0$ from 30~mT to 90~mT, the magnitudes of the profiles $B_{{\rm sc},y}(x)$, $B_{{\rm sc},x}(x)$ are increasing exactly three times -- see Eqs.~(\ref{eq:curr_elem}) and (\ref{eq:b_field}).

In the absence of the SC strip, the static magnetization is oriented out of plane due to the PMA in the Ga:YIG layer but the in-plane component $B_{{\rm sc},x}$ tilts the magnetization from the layer's normal; see Fig.\ref{fig:sc_stray_field}(c). Surprisingly, the angle between the magnetization and the applied field's direction increases with  increasing magnitude of the applied field. However, it is understandable if we keep in mind that the in-plane component of the stray field, which is responsible for the magnetization tilting, increases with the applied field: $B_{{\rm sc},x}\propto B_0$ -- see Fig.~\ref{fig:sc_stray_field}(b). The angle, at which magnetization is deflected from the out-of-plane direction $\theta_M(x)$ (Fig.\ref{fig:sc_stray_field}(c)), can be calculated by finding the minimum of the free energy density (\ref{eq:F}) for successive positions $x$. Our calculations have shown that the inhomogeneous exchange interaction caused by noncollinear magnetization texture does not significantly influence the value of the magnetization deviation angle. Therefore, to estimate angle $\theta_M$, we considered the free magnetic energy density of the film in an external magnetic field applied at an angle $\theta_B$, neglecting the exchange interaction:
\begin{equation}
  \begin{split}
    F(\theta) = &-B M_s \cos(\theta- \theta_B)\\ &+ \tfrac{1}{2}\mu_0 M_{\rm s} \cos^2(\theta) - K_{\rm u} \cos^2(\theta)
  \end{split}\label{eq:Energy_angl}  
\end{equation}
with angle $\theta$ the trial orientation of the magnetization for which we look for a minimum of $F(\theta)$ at every position $x$ independently. The angle $\theta=\theta_M$ corresponding to the minimum: $F_{\rm min}=F(\theta_M)$ determines the equilibrium orientation of the magnetization. The first term in Eq.~(\ref{eq:Energy_angl}) corresponds to the Zeeman interaction with total magnetic field, the second term describes the demagnetizing energy (shape anisotropy), and the third term corresponds to the energy related to the out-of-plane uniaxial anisotropy. It is worth noting that both the magnitude $B(x)=|\mathbf{B}_{0}+\mathbf{B}_{\rm sc}(x)|$ and the  angle $\theta_{B} (x) = {\rm arctan} [B_{{\rm sc},x}(x)/(B_0+B_{{\rm sc},y}(x))]$ of total static magnetic field are position-dependent.

The minimization of Eq.~(\ref{eq:Energy_angl}) results in an equation for the local magnetization angle $\theta_M (x)$ as a function of the $x$-coordinate. For relatively small values of the angles $\theta -\theta_B$ and $\theta$, the trigonometric functions in Eq.~(\ref{eq:Energy_angl}) can be expanded in a Taylor series that results in a polynomial equation for $\theta_M$:
\begin{equation}
  \begin{split}
    \frac{1}{\mu_0}B\left[(\theta_M -\theta_B)-\frac{1}{6}(\theta_M -\theta_B)^3\right]\\+\left(\frac{2K_{\rm u}}{M_{\rm s}}-M_{\rm s}\right)\left[\theta_M-\frac{2}{3}\theta^3\right]=0, 
  \end{split}\label{eq:eq_angl}  
\end{equation}
which is relatively easy to solve. For larger values of the applied field $B_0$ and related deflection angles, the equilibrium orientation $\theta_M(x)$ must be found numerically. The inset plot of Fig.~\ref{fig:sc_stray_field}(c) shows the evolution of the maximum angle deviation $\theta_{M,\rm{max}}$ with the value of the applied magnetic field $B_0$.

It should be noted that the formation of noncollinear magnetization texture allows the SW excitation by an alternating magnetic field applied along the external static field $\mathbf{B}_0$. 
In this case, the SW modes will only be excited in the regions below the edges of the SC strips, where the static magnetization is tilted. The texture in the FM layer can be modified not only by the strength of the external field $B_0$ but also by  geometrical parameters such as  size and aspect ratio $w/t$ of the SC strip and its separation $g$ from the FM layer.

\begin{figure}[b]
\includegraphics[width=1.0\columnwidth]{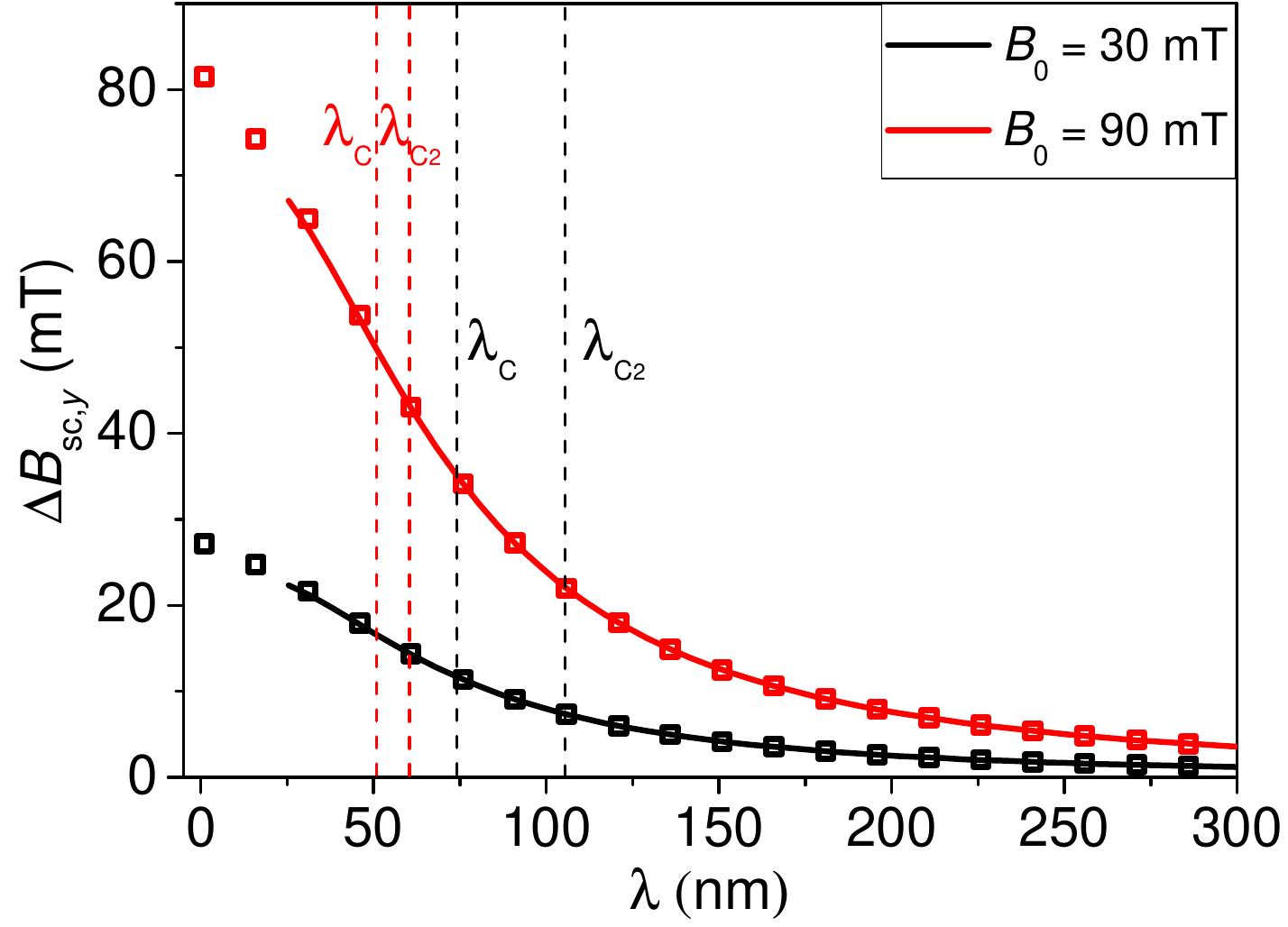}
\caption{The depth of the well of the SC field $\Delta B_{{\rm sc},y}$ (see Fig.~\ref{fig:sc_stray_field}(a)) as a function of the London penetration depth $\lambda$ calculated using Eqs.~(\ref{eq:b_field}) (solid lines) and numerical simulations (open squares). The black and red lines correspond to two different values of the external field $B_0=30$ mT and 90 mT, respectively. Dashed vertical lines show the critical values of the London penetration depth: $\lambda_C$ and $\lambda_{C2}$ corresponding to the case when the thermodynamic critical field is equal to external field: $B_{\rm C}=B_0$  and the transition to the normal state $B_{\rm C2}=B_0$,  for $B_0=30$ mT (black lines) and 90 mT (red lines), assuming the Ginzburg--Landau parameter $\kappa=1$.\label{fig:lambda}
}
\end{figure}

The London penetration depth $\lambda$ is a crucial parameter, that determines efficiency of the magnetic-field screening by the SC material. To understand how $\lambda$ influences on the condition for the SW localization, we have calculated the depth of the well $B_{{\rm sc},y}(x)$ (see Fig.~\ref{fig:sc_stray_field}(a)) as a function of $\lambda$ for the SC strip width $w=400$ nm and two selected values of the external field $B_0=30$~mT (black lines with open squares) and 90 mT (red line with open squares) -- see Fig.~\ref{fig:lambda}. Semi-analytical calculations performed with the help of Eqs.~(\ref{eq:b_field}) (solid lines) are in perfect agreement with the results of numerical simulations (square dots) for $\lambda>20$~nm, while for smaller values of $\lambda$, semi-analytical computations face difficulties. The dependencies $\Delta B_{{\rm sc},y}(\lambda)$ are presented in Fig.~\ref{fig:lambda}. It is evident that the decrease in $\Delta B_{{\rm sc},y}$ with increasing $\lambda$ is a manifestation of the gradual disappearance of SC properties, which is revealed as a weakening stray field that allows the external field to penetrate deeper into the SC strip. However, this picture becomes more complicated when we consider the more general theory of superconductivity described by the Ginzburg--Landau model that allows the phase transitions to mixed and normal phases at the critical fields $B_{\rm C1}$ and $B_{\rm C2}$, respectively, for type-II superconductors like Nb. Taking into account the dependencies of the critical fields on material parameters: penetration depth $\lambda$ and correlation length $\xi$, we should look for the critical values of the material parameters for which the Meissner state or superconductivity at all is destroyed. To estimate critical values of $\lambda$, we have used  the general relations between critical fields in a bulk superconductor \cite{Tinkham1996}: $B_{\rm C2}=\Phi_0 / (2\pi\xi^2)$ and $B_{\rm C2}/B_{\rm C}=\kappa\sqrt{2}$, $B_{\rm C1}/B_{\rm C}<1$ with $\Phi_0$ the magnetic flux quantum and $B_{\rm C}$ the thermodynamic critical field. Since giving an approximate analytical formula for $B_{\rm C1}$ is in principle possible for $\kappa\gg 1$ (for Nb: $\kappa=\tfrac{\lambda}{\xi}\approx 1$ \cite{McConville_1965}), we can only determine $\lambda_{\rm C}$ corresponding to the case $B_{\rm C}=B_0$, which is an upper estimate for $\lambda_{\rm C1}$ -- i.e. the critical London penetration depth at which $B_{\rm C1}=B_0$. For $B_0=30$~mT, $\lambda_{\rm C1}<\lambda_{\rm C}=74$~nm and $\lambda_{\rm C2}=105$~nm (black dashed lines in Fig.~\ref{fig:lambda}). This suggests the existence of the Meissner state and the validity of the  London's theory for $\lambda=\xi=50$~nm at the field $B_0=30$~mT.  However, for $B_0=90$~mT, we obtain $\lambda_{\rm C1}<\lambda_{\rm C}=51$~nm and $\lambda_{\rm C2}=60$~nm (red dashed lines in Fig.~\ref{fig:lambda}) which indicates the possible appearance of a mixed state with vortices. We confirmed this prediction by  FEM simulation of the Ginzburg--Landau model (see Appendix~\ref{app:stray_field_GL}
), where we numerically determined $B_{\rm C1}=53$~mT (32 mT) for the strip of the width 400~nm (800~nm) and thickness 100~nm, made of Nb where $\lambda=\xi= 50$~nm. Despite of the fact that the system under consideration is in a mixed state for the field $B_0=90$~mT, our results will be qualitatively valid for the case of other materials and structures that are able to sustain the Meissner state.

It is also worth noting that the London penetration depth increases with temperature. However, in helium temperatures, this change is less that 3.5\% for Nd, where the critical temperature is $T_{\rm C}=9.2$ K \cite{McConville_1965} -- see Appendix~\ref{app:temp_depend}
for more details.

\subsection{Spin-wave modes confined in the well of the magnetic field induced by a superconducting strip}

Figure~\ref{fig:modes} presents the SW profiles of the modes localized in the well of the effective field produced by the SC strip, for the external fields: $B_0=10$~mT (Fig.~\ref{fig:modes}(c)), 30~mT (Fig.~\ref{fig:modes}(a,d)) and 90~mT (Fig.~\ref{fig:modes}(b,e)). The profiles were computed numerically with the help of Mumax3 software (light red solid lines) and calculated using the semi-analytical model (red dashes lines). The blue lines show the FMR frequency of the homogeneous film at certain $B_0$ fields in the absence of the SC strip. The left and right columns of Fig.~\ref{fig:modes} correspond to the strip widths $w=400$~nm (Fig.~\ref{fig:modes}(a,b)) and 800~nm (Fig.~\ref{fig:modes}(c,d,e)), respectively. For larger values of the external field $B_0$, the well $B_{{\rm sc},y}(x)$ is deeper (see Fig.~\ref{fig:sc_stray_field}(a)) and can accommodate more SW modes. However, there is an upper limit for $B_0$ caused by the phase transition at $B_{\rm C1}$. Our model, based on London's theory, does not include the existence of a mixed state, which can appear in the considered range of the external field $B_0$. The presence of vortices will modify the stray field produced by the SC strip that will be important in the experimental implementation of the investigated system. However, the ultimate limit is given by the second critical field \cite{Williamson1970}.
On the other hand, the field $B_0=10$~mT is too small for the strip of width $400$~nm to create a well that is deep enough to confine SW modes. For the external fields $B_0=30$~mT (Fig.~\ref{fig:modes}(a)) or $90$~mT (Fig.~\ref{fig:modes}(b)) the well is sufficiently deep to bind one or two SW modes (solid and dashed lines) of the frequencies lower than the FMR frequency of the homogeneous film (horizontal blue line). Another strategy for SW binding is to widen the SC strip and thus widen the related well of the stray field.

\begin{figure}[h!]
\includegraphics[width=0.99\columnwidth]{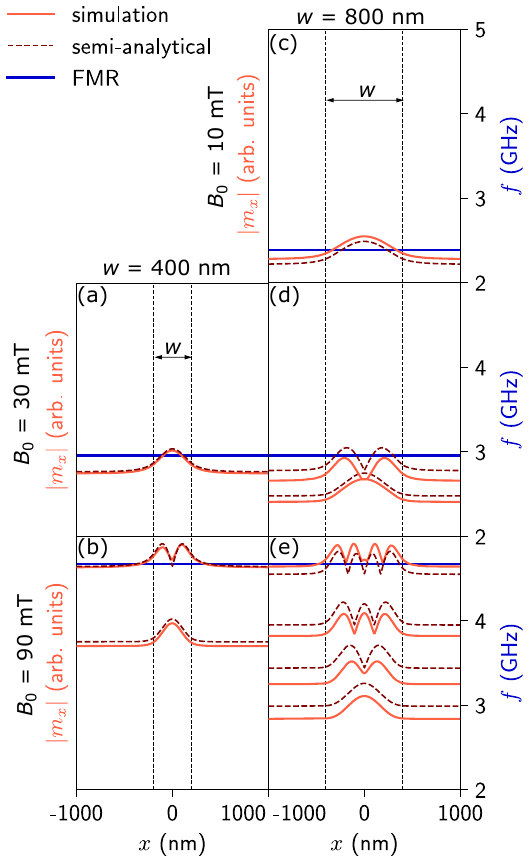}         
\caption{The SW profiles of the modes localized in the effective field well produced by the SC strip for the external field values $B_0=10$~mT (c), $30$~mT (a,d) and $90$~mT (b,e) obtained from micromagnetic simulations using the Mumax3 environment (light red solid lines) and the results of the semi-analytical model (red dashed lines). The blue lines show the FMR frequency of a homogeneous film for a given external magnetic field $B_0$ value in the absence of the SC strip. The left and right columns correspond to the SC widths $w=400$~nm and $800$~nm, respectively. Vertical black dashed lines indicate the position of the SC strip edges. All the plots share the same frequency and space scales and the amplitude of the modes is normalized to enhance  readability of the plot (the difference between the results obtained by the semi-analytical approach and micromagnetic simulations is shown by the offset in the plot). }\label{fig:modes}
\end{figure}

In Fig.~\ref{fig:modes}(c-e), for $w=800$ nm, we observe a larger number of localized SW modes, what is the result of two factors: (i) the frequency difference between SW-mode energy levels becomes smaller due to the widening of the well (the main factor), and (ii) the depth of the well is slightly increased with the widening of the SC strip  (e.g., for the $w=800$~nm $\Delta B_{{\rm sc},y}$ is about 1.4 times larger than for $w = 400$~nm). Thus, for $w = 800$~nm, there are one, two, and four localized modes for $B_0=10$~mT (Fig..~\ref{fig:modes}(c)), $30$~mT (Fig.~\ref{fig:modes}(d)), and $90$~mT (Fig.~\ref{fig:modes}(e)), respectively.
It should be noted that the approximation of the well $B_{{\rm sc},y}(x)$ by a parabolic well (\ref{eq:B_expansion}) is valid in the considered system only for a finite range of the widths $w$ (around $350$~nm), while for much smaller or larger widths, the terms higher than quadratic must be included in  expansion (\ref{eq:B_expansion}). For example, for $w=400$~nm and $800$~nm, the approximation of $B_{{\rm sc},y}$ requires including up to fourth-  and six-order terms, respectively. The semi-analytical results (dashed lines) are in relatively good agreement with the numerical solution of the LL equation obtained using the MuMax3 environment (solid lines). Because of the presence of higher-order terms in $B_{{\rm sc},y}$, the SW modes are not equidistant in contrast to the eigenfunctions of the quantum harmonic oscillator, i.e., Hermitian functions. However, our mode profiles can be approximated by Hermitian functions with fitting parameter $\beta_2$ in (\ref{eq:psi}). This trick allowed us to simplify the averaged demagnetizing field to a compact form (\ref{eq:h_dyn_aver}) (see Appendix~\ref{app:avg_hd}) and easily find the eigenfrequencies (\ref{eq:omega_diag}). 

\begin{figure}[!ht]
\includegraphics[width=0.99\columnwidth]{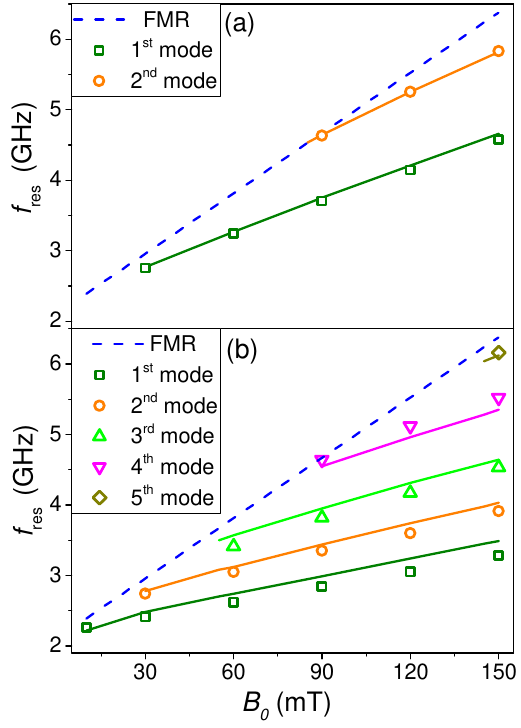}      
\caption{The dependence of the localized SW modes on the external magnetic field $B_0$ for the systems with SC strip widths $w=400$~nm (a) and $w=800$~nm (b). The solid lines and open squares correspond to the semi-analytical calculations and micromagnetic simulations respectively. Dashed lines show the FMR frequency of the uniformly-magnetized film.   
}\label{fig:field_dependence}
\end{figure}

The dependence of the frequencies of the localized SW modes on the external magnetic field is presented in Fig.~\ref{fig:field_dependence} for the SC strip widths $w=400$~nm (Fig.~\ref{fig:field_dependence}(a)) and $800$~nm (Fig.~\ref{fig:field_dependence}(b)). The results of the semi-analytical calculations and micromagnetic simulations are marked by solid lines and dots, respectively. As mentioned in the previous paragraph, the number of localized modes increases with increasing external field value and the width of the SC strip. In general, the results of the theoretical calculations and micromagnetic simulations are in good agreement. The discrepancies are larger for the strip width $w=800$~nm than for $400$~nm and they increase with increasing external magnetic field. This can be related to the simplifications underlying our theoretical model, according to which we have neglected the tangential component of the field produced by the SC strip $B_{{\rm sc},x}$, which slightly deflects the static magnetization from the normal to the FM layer and forms a noncollinear magnetization structure in the vicinity of the SC strip edges (see Fig.~\ref{fig:sc_stray_field}(c)). In our model, we have considered the uniform magnetization ground state where the FM layer is magnetized out of plane. This assumption is reasonable because the magnetization deviation angle $\theta_M$ is relatively small for a narrow strip, while for a wider strip, the deviation angle is larger and increases significantly for large external field values. To prove that this is the main reason for the discrepancies between the micromagnetic simulations and the semi-analytical model, we have performed micromagnetic simulations with $B_{{\rm sc},x} = 0$, which showed results much closer to the theoretical calculations. Small differences can also be caused by the use of the diagonal approximation in our theory. However, the diagonal elements of the dynamical demagnetization field are quite small, which means that nondiagonal elements can be really neglected because they are usually smaller than diagonal elements. 

\section{Conclusions}
We examined a superconductor--ferrimagnet hybrid planar nanostructure wherein a flat SC strip is situated above a uniform FM layer while the external field is applied out of plane. In this configuration, the flat SC strip, being in a Meissner state, produces a stray field that reduces the static effective field inside the thin FM layer placed below. When the FM layer is made of soft magnetic material with PMA, the SW modes can be confined within the region of lowered field (i.e., in the well of the internal field). These modes have frequencies lower than the FMR frequency of the FM layer, which explains their exponential decay outside the region of the well. By adjusting the value of the uniform external field, we can control the depth of the well and modify the number and frequencies of the confined SW modes. This outcome is difficult to achieve in the conventional magnonic nanostructures where the stray field, as a result of demagnetization, depends on the geometry and magnetic configuration. The first is determined at the fabrication stage and the latter can be modified by an external field, but its full control requires a strong field due to large demagnetization effects in planar structures.

Our work integrates the numerical and semi-analytical studies. The analytical approach is computationally undemanding and provides deep insight into the magnetization dynamics and the mechanism of SW localization in dipolarly-coupled superconductor--ferrimagnet systems. The presented ideas can be used to design superconductor--ferrimagnet hybrid devices for on-demand control of SW localization and propagation.

\begin{acknowledgements}
The authors would like to thank M. Silaev, O. Dobrovolskiy, O. Tartakivska and M. Zelent for fruitful discussions. The work was supported by the National Science Center, Poland, under Grants No. UMO-2019/35/D/ST3/03729, UMO-2021/43/I/ST3/00550, and UMO-2021/41/N/ST3/04478. 
The numerical simulations were performed at Poznań Supercomputing and Networking Center (Grant No. pl0095-01). K. Sobucki is a scholarship recipient of the Adam Mickiewicz University Foundation for the academic year 2023/2024.
\end{acknowledgements}

\section*{Data availability}
Data supporting this study are openly available from Zenodo:https://zenodo.org/records/10405886.

\appendix

\section{Stray field generated by magnetization texture in the FM layer -- micromagnetic studies}\label{app:micromag}
The stray field generated by SC eddy currents influences the FM layer, which is also manifested by the formation of a weak magnetization texture. We calculated the stray field produced by this texture to assess its impact on the SC strip. The calculations were done using micromagnetic simulations in the Mumax3 environment, with the external field $B_0=90$~mT applied along the $y$-axis. The stray field distributions are presented in Fig.~\ref{fig:fm_stray_field} for the strips of widths (a) $400$~nm and (b) $800$~nm. In both cases, the magnitude of the stray field is smaller than $1\%$ of the external field value, thus its influence on the SC system can be neglected and, therefore, the complicated self-consistent problem can be reduced to necessity to take into account only the impact of the SC strip on the FM layer. 

\begin{figure}[t]
    \centering
    \includegraphics{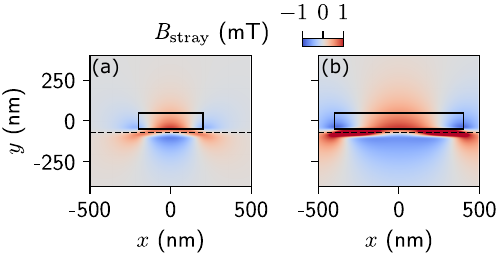}  
    \caption{The micromagnetic simulations of the stray field generated by the magnetization texture in the FM layer of  thickness $20$~nm. The texture is induced by  infinitely-long SC strips of thickness $100$~nm and widths (a) $400$~nm and (b) $800$~nm (separated from the FM film by the 10~nm air gap), placed in the external field $B_{0}~=~90$~mT applied in the $y-$direction. The black rectangle indicates the position of the SC strip and the dashed black lines marks the center of the FM film. In both cases: (a) and 
    (b), the magnitude of the stray field is less than $1\%$ of the external field value, so its influence on the system can be neglected. The material parameters for the FM film are the same as in the main text. The London penetration depth for the SC strip is $\lambda=50$~nm.}\label{fig:fm_stray_field}
\end{figure}

\section{Dispersion relation -- derivation}\label{app:disp}
The dispersion relation can be derived from Eqs.~(\ref{eq:lin_LL_op}) after introducing the average of the dipolar field (dipolar operator) in a way similar to the approach known in quantum mechanics. Let us express each SW mode by a function from the orthonormal base:  $m_{p,n}(x)\approx C_p M_{\rm s}\psi_n(Kx)$ as in Eq.~(\ref{eq:psi}). When we introduce the norm $\langle\psi^2_n\rangle:=\int_V dv |\psi_n(Kx)|^2\!=\!\frac{ab}{K}$, and then multiply both equations in (\ref{eq:lin_LL_op}) by $\psi^*_n(Kx)$ and integrate them over the whole volume of the magnetic system, we obtain: 
\begin{equation}
     \begin{split}
   \;\;i\omega\; C_x \langle\psi^2_n\rangle &=C_z \langle\psi_n|\Xi(x)|\psi_n\rangle-|\gamma|\mu_0 \langle h_{{\rm d},z} \rangle_n,\\
   -i\omega\; C_z \langle\psi^2_n\rangle&=C_x\langle\psi_n|\Xi(x)|\psi_n\rangle-|\gamma|\mu_0 \langle h_{{\rm d},x} \rangle_n,
   \end{split}\label{eq:lin_LL_2}  
\end{equation}
where $\langle h_{{\rm d},p} \rangle_n:=\int_Vdv\psi^*_n(Kx)h_{{\rm d},p}(x)$  for $p=\{x,y\}$ can be considered as an averaged component of the dynamic demagnetizing field. This definition is clear when we note that the field $h_{{\rm d},p,n}$ is the result of an action of the integral operator on $\psi_n(Kx)$ (strictly related to the magnetization $m_{p,n}(x)$) -- see Eq.~(\ref{eq:hd_gen}).   
Taking into account the fact that the $\psi_n(Kx)$ are eigenfunctions of the operator $\Xi(x)$ with the eigenvalues $\omega_{0,n}$, but not of the dynamical demagnetizing field operator, Eqs.~(\ref{eq:lin_LL_2}) can be rewritten as:
\begin{equation}
  \begin{split}
    \;\;i\omega\; C_x &=C_z\big(\omega_{0,n}-|\gamma|\mu_0 M_{\rm s}\langle \tilde{h}_{{\rm d},z}\rangle_n\big),\\
    -i\omega\; C_z &=C_x\big(\omega_{0,n}-|\gamma|\mu_0 M_{\rm s}\langle \tilde{h}_{{\rm d},x}\rangle_n\big),
  \end{split}\label{eq:lin_LL_3}  
\end{equation}
where $\langle\tilde{h}_{{\rm d},p}\rangle_n=1/(C_p M_{\rm s})\langle h_{{\rm d},p}\rangle$ is a dimensionless averaged field (see Appendix~\ref{app:avg_hd}). The homogeneous set of equations (\ref{eq:lin_LL_3}) gives nontrivial solution for the resonance frequency $\omega_n$ when
\begin{equation}
     \begin{split}
   \begin{vmatrix}
\;\;i\omega\; & \omega_{0,n}-|\gamma|\mu_0 M_{\rm s}\langle \tilde{h}_{{\rm d},z}\rangle_n \\
\omega_{0,n}-|\gamma|\mu_0 M_{\rm s}\langle \tilde{h}_{{\rm d},x}\rangle_n & -i\omega
\end{vmatrix}=0
   \end{split}\label{eq:det},  
\end{equation}
which leads to the dispersion relation shown in Eq.~(\ref{eq:omega_diag}).

\section{Averaged dipolar field -- derivation}\label{app:avg_hd}
For every $n^{\rm th}$ SW mode $\mathbf{m}_n(x)$ (confined in the well of the stray field), the dynamical demagnetizing field  $\mathbf{h}_{{\rm d},n}(x)$ (related to own dynamics of $\mathbf{m}_n(x)$) is given by Eq.~(\ref{eq:hd_gen}). The field $\mathbf{h}_{{\rm d},n}(x)$ can be considered as a result of the action of the integral operator on magnetization $\mathbf{m}_n$. When we approximate $\mathbf{m}_n$ by  function $\psi_n(Kx)$ from orthonormal set: $m_{p,n}(x)\approx C_p M_{\rm s}\psi_n(Kx)$ (see Eq.~(\ref{eq:psi})) and integrate $\mathbf{h}_{{\rm d},n(x)}$ over the whole volume with  $\psi_n(Kx)$, we obtain the average value of the dipolar field operator:
\begin{equation}
\begin{split}
   \langle h_{{\rm d},p}\rangle_n\!=\!-M_{\rm s}\sum_{q}C_q\int_V\int_{V'}\!dvdv'\;&\psi_n(Kx)\psi_n(Kx')\times\\ &\frac{d^2}{dp\,dq'}\frac{1}{4\pi|\mathbf{r}-\mathbf{r'}|}.\label{eq:dip_op_ave}
\end{split}
\end{equation}
Here the indices $p$ and $q$ run over the coordinates $x$ and $z$, and the derivative $\frac{d}{dq'}$ means $\frac{d}{dx'}$ (or~$\frac{d}{dz'}$) for $q=x$ (or~$q=z$). Each of the integrals in (\ref{eq:dip_op_ave}) is actually a threefold integral: $\int_{-\infty}^{\infty}dx\int_{y_0-a/2}^{y_0+a/2}dy\int_{-b/2}^{b/2}dz$ with $y_0=-(t/2+g+a/2)$. This problem can be simplified when we
express the function $\frac{1}{4\pi|\mathbf{r}-\mathbf{r'}|}$ in  Fourier space for the wave vector $\mathbf{k}=k_x \hat{\mathbf{x}}+k_z\hat{\mathbf{z}}$ \cite{Guslienko2010}, i.e.,
\begin{equation}
\begin{split}
  \frac{1}{4\pi|\mathbf{r}-\mathbf{r'}|}=\frac{1}{2\pi}\!\int_{-\infty}^{\infty}\int_{-\infty}^{\infty}\!dk_xdk_z \;\frac{1}{k}e^{-k|y-y'|}\times\\e^{ik_x(x-x')} e^{ik_z(z-z')},  
  \end{split}
\end{equation}
where $k=\sqrt{k_x^2+k_y^2}$, and then use the identities:

\begin{equation}
\begin{split}
    \int_{y_0-a/2}^{y_0+a/2}\int_{y_0-a/2}^{y_0+a/2}\!\!dydy'\;\frac{1}{k}e^{-k|y-y'|}=\frac{2a}{k}\left(\!1-\frac{1-e^{-ka}}{ka}\!\right),\\
    \int_{-b/2}^{b/2}\int_{-b/2}^{b/2}\!\!dzdz'e^{ik_z(z-z')}=2\pi b\;\delta(k_z), \;{\rm for}\; b\!\rightarrow\!\infty.
    \end{split}
\end{equation}
This allows us to express (\ref{eq:dip_op_ave}) in the form
\begin{equation}
\begin{split}
     \langle h_{{\rm d},p}\rangle_n\!=\!-M_{\rm s}\frac{ab}{K^2}\sum_{q}C_q\int dk_x\left( \int dk_z\frac{k_p k_q}{k^2}\times \right.\\
    \left. \left(1-\frac{1-e^{ka}}{ka}\right)\delta(k_z)\,\tilde{\psi}_n^2\!\left(\frac{k_x}{K}\right)\right).
    \end{split}\label{eq:dip_op_ave_2}
\end{equation}
One can note that $\langle h_{{\rm d},z}\rangle_n=0$ and $\langle h_{{\rm d},x}\rangle_n$ contains only the term for $p=x$. 
In the linear regime, the dynamic demagnetizing field scales linearly with the SW amplitude; therefore, it is useful to introduce the normalized dimensionless average field
\begin{equation}
     \langle \tilde{h}_{{\rm d},x}\rangle_n=\frac{1}{ C_x M_{\rm s}} \langle h_{{\rm d},x}\rangle_n,\label{eq:dip_op_norm}
\end{equation}
where $C_x M_{\rm s}$ is a SW amplitude in A/m. Finally, Eqs.~(\ref{eq:dip_op_ave_2}) and (\ref{eq:dip_op_norm}) lead to Eq.~(\ref{eq:h_dyn_aver}).

\section{ Stray field generated by the SC strip: the London {\em versus} Ginzburg--Landau model -- FEM studies}\label{app:stray_field_GL}

The Ginzburg--Landau (GL) theory is a generalization of the London theory of superconductivity. The use of the London theory in this paper is justified by its relative simplicity, which allows semi-analytical calculations. However, the London theory does not predict the existence of the critical fields $B_{\rm C1}$ and $B_{\rm C2}$, which are associated with the emergence of a mixed state (with magnetic vortices) and the complete disappearance of superconductivity, respectively. Therefore, we should investigate whether it is rational to consider the Meissner state in our system for the assumed value of the London penetration depth $\lambda= 50$~nm.

\begin{figure*}[!ht]
    \centering
    \includegraphics[width=0.95\linewidth]{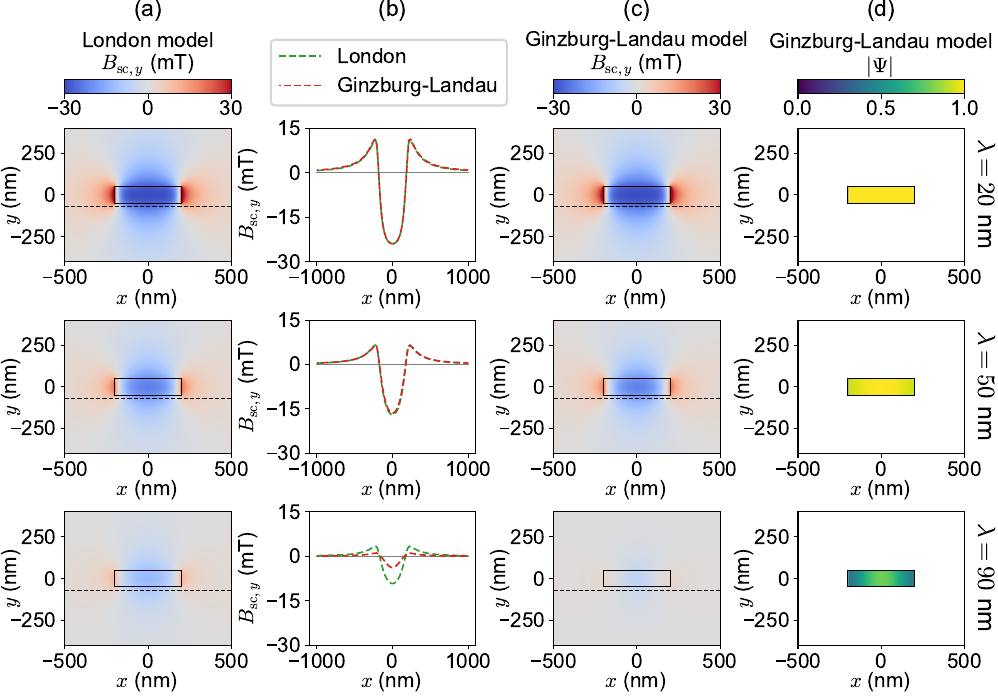}
    \caption{The FEM solutions of the GL and London equations for an infinitely-long SC strip of the thickness 100~nm and width 400~nm. The calculations were performed for different values of penetration depths $\lambda$ = 20, 50 and 90~nm. For simplicity, we assumed a correlation length $\xi=\lambda$, which leads to $\kappa=1$. The value of the external field $\mathbf{B}_0=B_0\hat{\mathbf{y}}$ has been set to 30~mT. 
    (a) London model: the $y$-component of the stray field $B_{{\rm sc},y}$ on the $xy$-plane inside the SC strip and in the surrounding non-magnetic and non-conducting domain. The dashed black line indicates the center of the FM layer.
    (b) The $y$-component of the stray field $B_{{\rm sc},y}(x)$ obtained from the London and GL models in the center of the FM layer ($y=-70$ nm).
    (c) GL model: the $B_{{\rm sc},y}$ on the $xy$-plane inside the SC strip and in the surrounding non-magnetic and non-conducting domain. The dashed black line indicates the center of the FM layer.
    (d) GL model: the modulus of order parameter $|\Psi|$ in the SC domain.
    }
\label{fig:londonvsgl}
\end{figure*}

We used COMSOL Multiphysics to solve dimensionless time-dependent GL equations in the SC domain \cite{Oripov2020, alstrom2011magnetic,Gulian2020}:
\begin{equation}
\begin{split}
&\eta\frac{\partial \Psi}{\partial t}=-\Big(
\frac{i}{\kappa}\nabla + \mathbf{A} \Big)^2 \Psi +
\left(1-\rvert \Psi \lvert^2 \right) \Psi, \\ &
\sigma \frac{\partial \mathbf{A}}{\partial t}=
\frac{1}{2 i\kappa} \left(\Psi^* \nabla \Psi - \Psi \nabla \Psi^* \right) - \rvert \Psi \lvert ^2 \mathbf{A} - \nabla \times \nabla \times \mathbf{A}.
\end{split}\label{eq:GL}
\end{equation}
Here $\Psi=\Psi(\mathbf{r}, t)$ is the order parameter and  $\sigma$ represents the electric conductivity of the normal (non-SC) state, while $\eta$ denotes the normalized friction coefficient. Since we have focused on the steady state of the system ($t\rightarrow\infty$) rather than its dynamics, the choice of specific values of $\sigma$ and $\eta$ for the dimensionless GL equation is not relevant once we are able to reach the same steady state. Outside the SC strip (in the vacuum) $\Psi=0$, $\sigma=0$, and the vector potential was determined from the equation
\begin{equation}
   \nabla \times \nabla \times \mathbf{A} =0.
    \label{eq:amper_gl}
\end{equation}
To ensure that the GL equations are well defined, the gauge must be fixed. We have chosen the gauge with zero electric potential, which turns out to be the most convenient choice for Eqs.~(\ref{eq:GL}). 

We set the boundary conditions at the interface between SC and vacuum as $\nabla \Psi\cdot\mathbf{n}=0 $ and $\mathbf{A}\cdot\mathbf{n}=0$ \cite{Oripov2020}. The vacuum domain is large in size but finite. We assumed that the SC stray field on its boundaries equals zero and total field $\mathbf{B}= B_0\hat{\mathbf{y}}$, which leads to the boundary condition $\nabla \times \mathbf{A} = B_0\hat{\mathbf{y}}$ \cite{Oripov2020}. We solved Eqs.~(\ref{eq:GL}) and~(\ref{eq:amper_gl}) using 2D model, i.e. assuming that $\mathbf{A}(x,y)$ and $\Psi(x,y)$ are homogeneous along the strip. The transitions from dimensionless to realistic quantities are described by the transformations: $(x,y,z)\rightarrow(x/\lambda, y/\lambda, z/\lambda)$, $t\rightarrow t/(\mu_0\lambda\sigma)$, and $\mathbf{A}\rightarrow \frac{2e \xi}{\hbar}\mathbf{A}$.

In this paper, we consider a 2D system and assume its homogeneity in the $z$-direction. This assumption is not valid in the mixed state with the field applied in the $y$-direction, where the induced vortices disturb the mentioned homogeneity. For this reason, the first critical field must be estimated numerically using a 3D model. This gives us the range of the applied field in which we are still in the Meissner state for a given value of $\lambda$ (assuming that we are working with a $\kappa\approx1$ superconductor), and we can investigate the system using London's model.

\subsection*{Two-dimenissional simulations}

We start with 2D simulations based on the GL model, which should give an insight into the reduction of the stray field associated with the change in the density of the Cooper pairs, which is not included in the London model.

    Figure~\ref{fig:londonvsgl} shows the difference between the 2D GL and London models for fixed external field $B_0=30$ mT and three selected values of penetration depth $\lambda=20$, 50, 90 nm. In our studies, we fixed the GL parameter $\kappa=\lambda/\xi=1$, which is equivalent to $\xi=\lambda$ (this choice is in the range of $\kappa$ for bulk Nb). For small values of $\lambda$ (and $\xi$) both models give consistent results. For $\lambda=20$ nm, the difference between the models is negligible. For $\lambda=50$ nm, used in the manuscript for the studies of confined SW modes, the differences in the stray field (calculated from the GL and London models) are still barely distinguishable -- see the middle plot of Fig.~\ref{fig:londonvsgl}(b). For $\lambda=50$ nm, $|\Psi|\approx 1$ in the whole volume of the SC strip except for a small reduction near its lateral edges (see the middle plot of Fig.~\ref{fig:londonvsgl}(c)), which means that we are still in the Meissner state, i.e., the external field is still below the first critical field: $B_0<B_{\rm C1}$. For $\lambda=90$ nm, the discrepancy between the London and GL models becomes significant. The stray field produced by the SC is about 2 times smaller in the GL model; see the bottom plot of Fig.~\ref{fig:londonvsgl}(b). At the same time, $|\Psi|$ is strongly reduced; see the bottom plot of Fig.~\ref{fig:londonvsgl}(d).

\begin{figure}[t]
    \centering
    \includegraphics{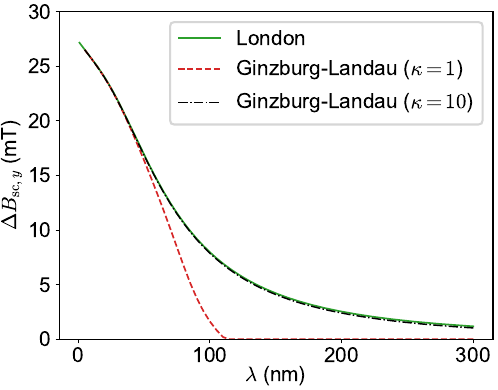}
    \caption{The depth of the well of the SC field $\Delta B_{{\rm sc},y}$ (at the external field $B_0=30$ mT) as a function of the penetration length $\lambda$ calculated using the London model (solid green line) and GL model for $\kappa=1$ (dashed red line) and $\kappa=10$ (dash-dot black line).
    }
\label{fig:lambda_GL_L}
\end{figure}

Figure~\ref{fig:lambda_GL_L} is analogical to Fig.~\ref{fig:lambda} in the main part of the manuscript. It shows the depth of the well of the SC field for the London and the GL models at 30~mT. For comparison, we used two values of $\kappa$: 1 and 10. Up to $\lambda\approx 50$ nm, the well depth for all  systems is almost the same. For $\lambda> 50$ nm, it starts to decrease much stronger for the GL model with $\kappa=1$, reaching zero at the critical value of $\lambda=\xi= 115$ nm. Such an effect is understandable if we note that, as the second critical magnetic field $B_{\rm C2}\propto 1/\xi^2$, the increase in $\lambda$ (for constant $\kappa$) leads to a reduction in the second critical field $B_{\rm C2}$. In the point of the second critical field, the order parameter goes to zero, leading to the disappearance of superconductivity. Such a phase transition is not observed in the London model, which lacks any kind of critical field. Interestingly, if we increase $\kappa$ to 10, the results are in agreement with the London model, showing that, for large values of $\kappa$, it gives reliable results for a wide range of $\lambda$.

\subsection*{Three-dimensional simulations}

\begin{figure}[b]
    \centering
    \includegraphics[width=0.9\linewidth]{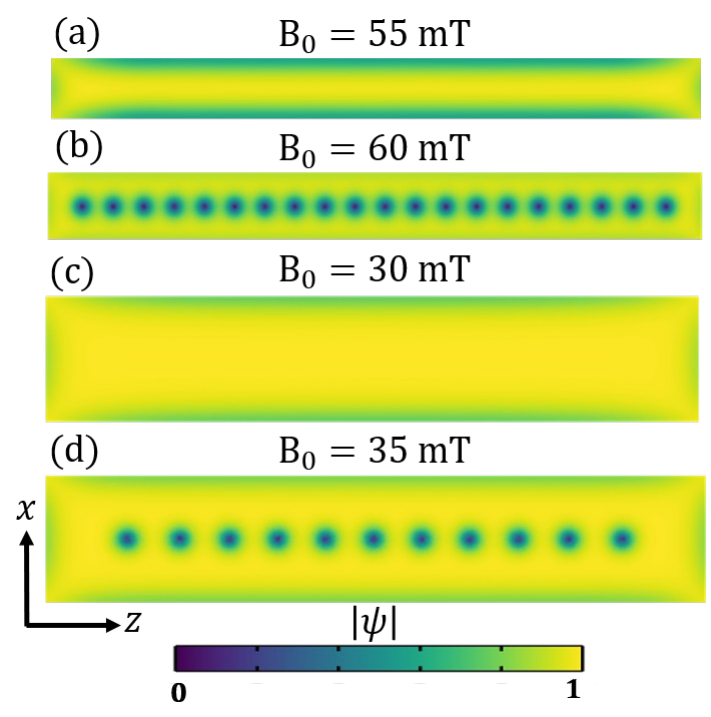}
    \caption{FEM simulations of the GL model of the nucleation of magnetic vortices in the finite SC strip  length  4000~nm and thickness 100~nm. We investigated strips of two different widths: (a), (b) 400~nm and (c), (d) 800~nm. For the narrower (wider) strip, the transition to a mixed state is observed at $B_{\rm C1}\approx$~53~mT ($B_{\rm C1}\approx$~32~mT). The modulus of the order parameter $|\Psi|$ is plotted in the $x-z$ plane in the centers of the strips ($y=0$), for the external field value slightly below (a), (c) and slightly above (b), (d) $B_{\rm C1}$. We assume that $\lambda=\xi=50$~nm.}
\label{fig:Bc1-GL}
\end{figure}


To observe the nucleation of vortices, we solved the GL equations using FEM simulations for the SC strips of finite length (4000 nm), which differ in width (400 nm and 800 nm) and have the same thickness (100 nm). Fig.~\ref{fig:Bc1-GL} presents the distribution of the modulus of the order parameter $|\Psi|$ along the strips and across their widths (i.e. in the $x-z$ plane). We can see the transition from the Meissner state [Fig.~\ref{fig:Bc1-GL}(a,c)] to mixed state [Fig.~\ref{fig:Bc1-GL}(b,d)], where a single row of Abrikosov vortices appears in the center of the nanowire at higher fields. The numerically estimated value of the first critical field $B_{\rm C1}$ is larger for the narrower strip (53 mT) than for the wider one (32 mT).

The appearance of vortices allows the external field to penetrate into the superconductor, which significantly reduces the stray field produced by the SC strip. It makes the SW binding much more difficult. Since vortices generate inhomogeneity along the strip axis, the problem cannot be treated as quasi-one-dimensional for spin waves, which may be the subject of a new study.

\section{Temperature and field dependence of the London penetration depth}\label{app:temp_depend}

\begin{figure}[b]
    \centering
    \includegraphics[width=1.0\linewidth]{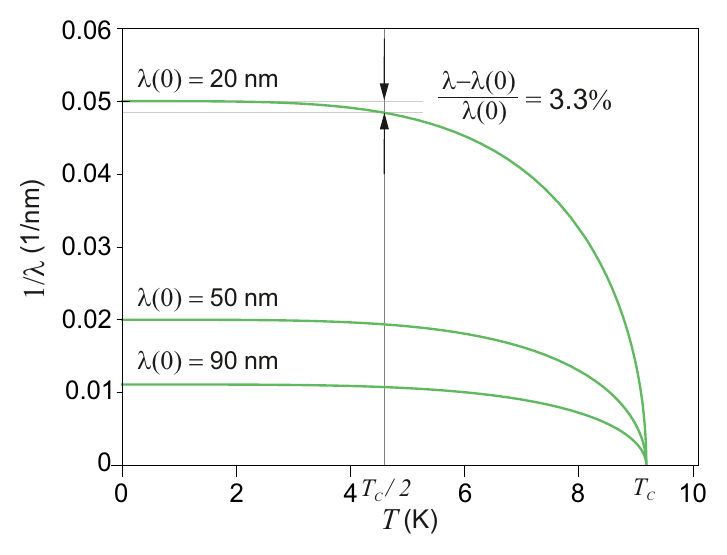}
    \caption{The temperature dependence of the inverse of the London penetration depth $1/\lambda(T)$. We used Eq.~(\ref{eq:lambdaT}) for three (considered in Fig.~\ref{fig:londonvsgl}) values of London penetration depths at $T=0$, $\lambda(0)=$ 20, 50, 90 nm, and for fixed critical temperature $T_C=9.2$ K, characteristic for Nb. In the wide temperature range, $T<T_C/2$, $\lambda(T)$ changes less than 3.3\%.}
\label{fig:lambda_temp}
\end{figure}

For practical reasons, it is also worth  studying how the London penetration depth changes with temperature $T$ and external field $B_0$. For a rough estimation of the $\lambda(T)$ dependence, the general formula based on a two-fluid model \cite{Tinkham1996,Bardeen1954} can be used:
\begin{equation}
    \lambda(T)=\lambda(0)\frac{1}{1-\left(\frac{T}{T_C}\right)^4},\label{eq:lambdaT}
\end{equation}
In the wide range of temperatures $0<T<T_C/2$, the relative change $(\lambda(T)-\lambda(0))/\lambda(0)$ is about 3.3\% -- see Fig.~\ref{fig:lambda_temp}. It is worth noting that for Nb ($T_C=9.2 K$), $T_C/2$ is close to the helium boiling temperature. 

The London penetration depth depends weakly on the applied field $B_0$ for low temperature $T<T_C/2$. For the most of the superconductors \cite{Pippard1950,Sonier1997}, the dependence $\lambda(B_0)$ exhibits an slight increase, quadratic in $B_0$, which can be ignored in our study.



%

\end{document}